\documentclass[conference]{IEEEtran}

\usepackage[super]{nth}
\usepackage{orcidlink}
\usepackage{ulem}
\usepackage{listings}
\usepackage{anyfontsize}
\definecolor{kento}{RGB}{255, 111, 97}

\def\BibTeX{{\rm B\kern-.05em{\sc i\kern-.025em b}\kern-.08em
    T\kern-.1667em\lower.7ex\hbox{E}\kern-.125emX}}
\begin{document}

\title{Understanding Power Consumption Metric on Heterogeneous Memory Systems}


\author{\IEEEauthorblockN{Andrès {Rubio Proaño}\orcidlink{0000-0003-1270-9140}}
\IEEEauthorblockA{RIKEN R-CCS\\
Kobe, Japan\\
andres.rubioproano@riken.jp}
\and
\IEEEauthorblockN{Kento {Sato}\orcidlink{0000-0001-7850-2121}}
\IEEEauthorblockA{RIKEN R-CCS\\
Kobe, Japan\\
kento.sato@riken.jp}
}


\maketitle

\begin{abstract}

    Contemporary memory systems contain a variety of memory types, each possessing distinct characteristics. This trend empowers applications to opt for memory types aligning with developer's desired behavior. As a result, developers gain flexibility to tailor their applications to specific needs, factoring in attributes like latency, bandwidth, and power consumption.
Our research centers on the aspect of power consumption within memory systems. We introduce an approach that equips developers with comprehensive insights into the power consumption of individual memory types. Additionally, we propose an ordered hierarchy of memory types. 
    Through this methodology, developers can make informed decisions for efficient memory usage aligned with their unique requirements.%This ordering proves invaluable when optimizing power savings or striking a balance between power consumption and application performance. 

\end{abstract}

\begin{IEEEkeywords}
power consumption; nvm; heterogeneous memory

\end{IEEEkeywords}

% \fi
\IEEEpeerreviewmaketitle

\section{Introduction}

%\reviewerone[]{7.The introduction is missing a discussion of the contributions. It goes straight from explaining the motivation to outlining the paper structure.}

High-performance computing (HPC) plays a crucial role in addressing a wide range of complex scientific challenges by utilizing advanced models and simulations. %The hardware components powering HPC systems consist of central processing units (CPUs), memory, storage, and networking devices, in addition to specialized accelerators optimized for various workloads. These components work in tandem to deliver exceptional computing capabilities, enabling researchers and scientists to push the boundaries of scientific discovery and innovation.

\begin{figure}[!htb]
  \center{\includegraphics[width=0.8\linewidth]
  {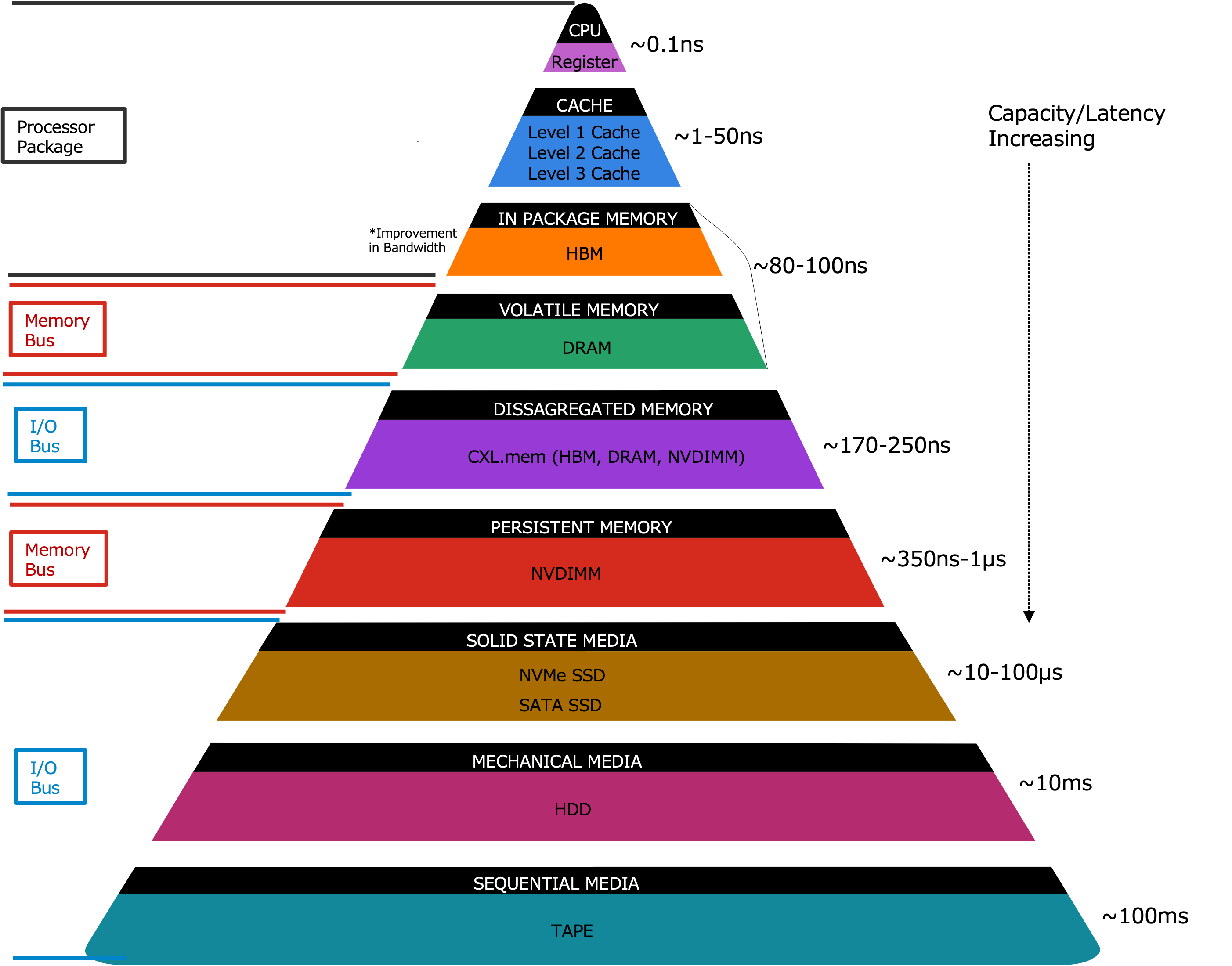}}
  \caption{\label{fig:memcontinuum} Memory-Storage Continuum.}
\end{figure}

Over the years, the architecture of supercomputers has undergone significant evolution to meet diverse computing requirements, including enhancing application performance and optimizing power utilization. A critical aspect of this evolution lies in the memory system, which has evolved into a heterogeneous structure with multiple levels or hierarchies, incorporating various technologies. In Figure~\ref{fig:memcontinuum}, the memory-storage continuum illustrates the primary technologies associated with each hierarchy level. 
%This continuum showcases the diverse range of memory and storage technologies used, each offering specific advantages in terms of speed, capacity, and power. This heterogeneity allows supercomputers to efficiently handle different types of workloads, ranging from data-intensive tasks to computation-heavy simulations, thereby enabling more flexible and efficient computing solutions. The ongoing advancements in memory technologies continue to drive the capabilities of modern supercomputers, empowering them to tackle ever more complex scientific and computational challenges.

%Heterogeneity in computing systems extends beyond processing units like CPUs and GPUs; it has also become prominent in memory systems. %A notable example include the Intel Xeon, Cascade Lake, that uses DRAM and non-volatile memory (NVM)~\cite{arafa2019cascade}.%Intel Xeon Phi, Knight Landing (KNL)~\cite{zhao2016estimating}, which uses dynamic random access memory (DRAM) and high bandwidth memory (HBM). %Also, we can see it in the Intel Xeon, Cascade Lake, that uses DRAM and non-volatile memory (NVM)~\cite{arafa2019cascade}. Additionally, Intel Xeon Sapphire Rapids introduces support for Compute Express Link (CXL) technology, enabling integration of DRAM, HBM, and NVM~\cite{cxl1}. 

%This increasing 
Heterogeneity within memory systems is driven by the diverse nature of applications, each exhibiting affinities towards specific memory devices. By leveraging these different memory technologies, improvements in latency, bandwidth, and memory power consumption behaviors can be achieved.%, further enhancing overall system performance. %As a result, modern computing architectures are becoming more adaptable and efficient, allowing them to cater to a wide array of workloads with varying demands. 
%Embracing memory system heterogeneity unlocks new avenues for addressing complex computational tasks and optimizing the overall performance of supercomputing and high-performance computing systems.

Given the above, this paper presents a methodology that allows expanding the knowledge of the memory system in terms of power, offering the possibility of obtaining an ordering of memories given a heterogeneous memory system when using different applications.

The paper structure considers first a Section~\ref{sec:stateofart} that provides an overview of the pertinent memory devices and their integration into Heterogeneous Memory Systems(HMS). In Section~\ref{sec:motivation}, we elucidate the driving factors behind this study. Section~\ref{sec:methodology} outlines our devised methodology for assessing application power consumption within HMS. Subsequently, Section~\ref{sec:evaluation} employs a suite of memory-bound applications to analyze memory power consumption within a heterogeneous memory system. The paper concludes by presenting future prospects and conclusions.
\section{State of the Art}
\label{sec:stateofart}

The acceleration in computation performance has been increasing enormously. However, memory systems have not been able to keep up with it. %This has led to a memory wall that various vendors and technologies are attempting to overcome. %The result of this search has increased the complexity of memory systems in terms of hierarchies, heterogeneity, power consumption, and management~\cite{wulf95}

Here we present some of the most relevant memory technologies.% involved in heterogeneous memory system.
\subsection{HBM}

%KNL, introduced several years ago, features on-package memory based on multi-channel dynamic random access memory (MCDRAM) technology. In KNL, this type of memory offers higher bandwidth compared to DRAM, which also coexists in these machines. 
Nowadays, High Bandwidth Memory (HBM) modules offers higher bandwidth and have becomed a part of other high-performance computing systems, including high-end GPUs and nodes equipped with A64FX processors. %Intel has also joined the fray, with the Intel Sapphire Rapids~\cite{sapphire} platform supporting HBM. This not only increases the available bandwidth significantly but also addresses a major limitation of KNL, which was its capacity constraint. In fact, it enables pure HBM configurations with larger problem sizes.

\subsection{NVM}

NVM, %enabled by Intel Optane Data Center Persistent Memory, 
is a type of memory that entered the market as a result of Intel's decision to bridge the growing gap between DRAM and NAND SSD. Their flexibility enables them to coexist with DRAM, fostering a heterogeneous environment that harmonizes the strengths of different memory technologies.%It does not adhere to any JEDEC standard~\cite{lee2020}; however, it is conceptually similar to NVDIMM-P.
%The introduction of Optane NVM, or simply NVM, seamlessly integrates into the second generation of Intel Xeon Scalable processors. In this mutually beneficial arrangement, NVM and DRAM coexist, capitalizing on their respective strengths to create a dynamic, heterogeneous memory environment. This nuanced integration gives rise to a range of compelling features that redefine memory paradigms. Notable attributes include terabyte-scale capacities, latencies on the order of hundreds of nanoseconds, and a remarkable capacity for data persistence.
%These memory technologies have found utility in dual roles, serving as both storage and memory solutions. .  %This synergy optimally addresses the diverse demands of modern computing workloads. In~\cite{goglinoportuni}, we can observe the details of how DRAM and NVM have coexisted, including using DRAM as a cache, using it as extra NUMA nodes, and using it as storage.

\subsection{CXL.mem}

CXL, which stands for Compute Express Link, %represents a pivotal advancement that transcends the conventional boundaries of memory itself. 
instead of being a distinct form of memory, CXL is, in fact, an industry-endorsed Cache-Coherent Interconnect meticulously engineered to seamlessly integrate processors, memory, and accelerators. This standard is underpinned by an innovative protocol known as CXL.mem~\cite{sharma2023introduction}.%, a foundational framework that facilitates access to a diverse array of memory types that can be harnessed through this interconnect.

%At the core of the CXL paradigm is the transformative ability to enhance memory performance and accessibility. This is achieved by enabling direct access to various memory types—ranging from traditional DRAM to cutting-edge technologies like HBM and NVM—all interconnected through CXL. This orchestration serves as a conduit for channeling greater bandwidth and capacity directly into processors, unshackling them from the constraints of the primary main memory. Intel's Sapphire Rapids architecture stands as a tangible testament to the adoption of CXL~\cite{sapphire}. This architecture underscores the industry's recognition of CXL's transformative potential, paving the way for a future where memory technologies are seamlessly interconnected and poised to unlock new thresholds of performance and capacity.

\subsection{DRAM}

%The evolution of DRAM throughout the years has been nothing short of remarkable. %It has consistently maintained its role as the principal main memory component in numerous computing systems. 
A notable contrast to traditional DRAM DIMMs is the use of LPDDR (Low-Power Double Data Rate) memory in NVIDIA Grace CPUs~\cite{lpddr}. %While traditional DRAM DIMMs have been the standard for main memory in computing systems for years, LPDDR brings a different approach to memory design, primarily focusing on power efficiency. In the context of heterogeneous memory systems, DRAM remains a unifying element, seamlessly interfacing with other advanced memory technologies such as HBM and NVM.
However, there are still gaps in the memory-storage continuum. The need for these gaps depends on the requirements of applications to reach new levels of performance and power efficiency.% Therefore, it can be predicted that new technologies will emerge, either to improve or replace the current ones, and to fill the gaps that remain unfilled.

\section{Related Work}

In earlier days, when memory systems exhibited homogeneity, there existed a sense of implicit assumptions regarding several factors, such as memory power. However, the landscape has transformed with the advent of heterogeneous memory systems.

Among these changes, several studies aimed to better handle diverse memory systems. For instance, in a paper referenced as~\cite{cantalupo2015memkind}, researchers explored ways to manage memory in heterogeneous systems. %They specifically looked at memory management in cases like MCDRAM and DRAM on the KNL platform. %This work showcases efforts to adapt memory management techniques to suit the unique characteristics of different memory types within heterogeneous systems.

In~\cite{ilsche2019power} we can observe tecniques for power efficienct computing in HPC systems by presenting measurement solutions that consider scalability, resolution and accuracy in power consumption measurements. %This approach considers external devices like specialized power-meter that in principle can give accurate values but is hard to implement.

%In studies like the one presented in~\cite{paul123}, a comprehensive exploration of power consumption is evident, with a primary emphasis on scrutinizing power usage from the processing angle. This research delves into the broader realm of power dynamics, offering insights into how power is consumed in relation to the processing elements of the system. While addressing a wider spectrum of power considerations, this work provides a valuable perspective on power management strategies that influence the overall efficiency and performance of computing systems.

In~\cite{rapl}, we can observe how authors give relevance to the memory power consumption metric. In fact, some old counters were used to understand power in memory; however, until the release of this paper, there was no way to differentiate the power consumption of each kind of memory.
\section{Motivation}
\label{sec:motivation}

High-Performance Computing (HPC) systems have witnessed remarkable advancements in recent years, enabling researchers and scientists to tackle increasingly complex and data-intensive computational problems. A significant driver of this progress lies in the utilization of heterogeneous memory systems, which combine diverse memory technologies such as dynamic random access memory (DRAM), high bandwidth memory (HBM), and non-volatile memory (NVM). %These memory systems offer unprecedented flexibility, allowing HPC architectures to match the specific demands of various workloads and optimize overall system performance.

However, with the proliferation of diverse memory devices, a new challenge arises: power consumption optimization. %Power is a critical concern in modern HPC systems due to the soaring power demands associated with processing large-scale simulations and data analysis tasks. 
As the scale and complexity of scientific simulations and data-intensive applications continue to expand, the power consumed by the memory subsystem becomes a significant portion of the total power consumption in the system~\cite{sarood2013}. %Therefore, optimizing power consumption within heterogeneous memory systems becomes paramount to curbing the soaring power costs associated with large-scale computing facilities.

Given this, there is a critical need to provide a comprehensive ranking or ordering of memory devices based on their power consumption characteristics. %By understanding the power consumption behaviors of different memory technologies under diverse workloads, HPC system designers and administrators can make informed decisions to enhance power consumption, reduce operational costs, and minimize the carbon footprint of supercomputing centers.

%Also, in the context of power capping, it is essential to provide a comprehensive ordering of memory devices based on their power consumption characteristics. This mechanism is implemented in modern HPC centers to limit the maximum power consumption of computing systems. As supercomputers continue to scale in size and complexity, managing power consumption becomes a fundamental challenge, necessitating the integration of power capping strategies to maintain stable operations and prevent potential thermal issues. With the integration of heterogeneous memory systems in HPC architectures, power capping becomes even more crucial, as different memory devices exhibit varying power consumption profiles under different workloads. In this context, understanding the power consumption behaviors of individual memory components is vital for devising efficient power capping policies that strike the optimal balance between performance and power restrictions.

%As demonstrated by Sapphire Rapids, the adoption of a heterogeneous memory system is a viable choice for future systems. In the current landscape, the trend toward heterogeneous memory systems entails the development of techniques that enable us to harness these diverse memory resources effectively. This pursuit encompasses a comprehensive grasp of memory power consumption. %This awareness is crucial, considering that contemporary supercomputer nodes must navigate scenarios where power availability is limited.

The power provisioning approach, as exemplified in~\cite{arima2022convergence}, demands the optimal allocation of each component within the High-Performance Computing (HPC) ecosystem, aligned with specific power requirements. Thus, the development of a pertinent strategy becomes pivotal, allowing us to establish a strategic ordering among memory devices. This ordering, in turn, contributes to effective power provisioning and allocation.
\section{Methodology}
\label{sec:methodology}

In the present day, applications are capable of generating diverse workloads characterized by unique behaviors, which can respond differently within a given HMS. Additionally, the proliferation of various HMS configurations has introduced challenges for developers.%, particularly when they lack direct access to these systems. In such scenarios, developers may need to modify or code applications to effectively accommodate specific memory metrics, such as power consumption, and optimize their performance accordingly.

Here, we present the methodology employed for characterizing the power consumption metric within a HMS. We commence by detailing the process of configuring and exposing our memory system. Subsequently, we expound upon the selection of benchmarks and applications.%, elucidating the procedures we adopted to conduct our tests across distinct memory types. 
Additionally, we introduce the tools to extract the requisite performance counters. Lastly, we use the information and undertake a comprehensive comparison of power performance variations across the various memory kinds.

\subsection{Identifying Heterogeneous Memory Systems}

The task of identifying memory system targets holds significant importance for both developers and applications. Particularly with the advent of HMS that incorporate multiple memory types, each characterized by distinct properties like bandwidth, latency, and power consumption.%, as it enables them to differentiate between various memory types.%This task's significance has escalated over the years, 

%In earlier times when systems had a single type of memory, the need to identify memory system kind was not as imperative. However, with the emergence of Non-Uniform Memory Access (NUMA) systems, which demanded consideration of locality, the task to identify the most suitable memory given locality became essential. In today's context of HMSs, the accurate identification and exposure of memory targets hold paramount importance. This is because applications can exhibit diverse behaviors and performance outcomes depending on the specific memory types employed.

A tool that greatly facilitates this task is $\tt{hwloc}$~\cite{goglin0}. This tool streamlines the discovery of hardware topologies in HPC platforms. %By modeling multi-level memory architectures and exposing memory devices connected to CPUs, $\tt{hwloc}$ offers developers and applications a more accessible means to discern the types of memory available within a system.

%\begin{figure}[!htb] 
%    \center{\includegraphics[width=0.9\linewidth]
%    {img/lstoposysram.png}}
%    \caption{Output of $\tt{hwloc}$'s lstopo tool on an Intel Xeon Cascade Lake with 2 sockets. Each CPU has local DRAM and NVM.}
%    \label{fig:timberwolf00}
% \end{figure}

It is crucial to highlight that NVM possesses the flexibility for two distinct applications: serving as storage or functioning as a memory device. In our study, our focus lies on utilizing NVM as a memory device to establish a HMS that incorporates both NVM and DRAM. %To achieve this, we employed the $\tt{daxctl}$ management utility command, which enabled us to expose memory devices as additional NUMA nodes.% In order to show NVM as extra numa nodes the following command was perform:

%\begin{lstlisting}[language=bash, basicstyle=\scriptsize]
  
%  $ daxctl reconfigure-device --mode=system-ram dax0.0
%\end{lstlisting}

The configuration of our test machine 
%is depicted in Figure~\ref{fig:timberwolf00}
 features two sockets housing second-generation Intel Xeon Scalable processors and distinct types of memory local to each socket. This configuration results in a total of four NUMA nodes.

%Subsequent to the exposure of our memory system, a critical step involves characterizing it in relation to performance metrics \cite{leon}. %This characterization serves the purpose of establishing the performance limitations of our memory device.

\begin{table}[!htb]
  \begin{center} 
  \caption{Power consumption over different applications in a HMS \\ using 18 threads and a big problem size   }
  \label{tab:powerbeha}
  \fontsize{6pt}{6pt}\selectfont
  \begin{tabular}{|c|c|c|}
    \hline
    \textbf{Application.size} & \textbf{\begin{tabular}[c]{@{}c@{}}DRAM Power \\ Consumption {[}Watts{]}\end{tabular}} & \textbf{\begin{tabular}[c]{@{}c@{}}NVM Power\\ Consumption {[}Watts{]}\end{tabular}} \\ \hline
    BT.A                  & 86.91                                                                                  & 74.73                                                                                \\ \hline
    CG.W                   & 87.5                                                                                   & 77.71                                                                                \\ \hline
    FT.A                   & 99.65                                                                                  & 72.91                                                                                \\ \hline
    IS.B                   & 97.41                                                                                  & 74.39                                                                                \\ \hline
    MG.B                   & 90.36                                                                                  & 75.21                                                                                \\ \hline
    hpccg.400                & 115.15                                                                                 & 78.16                                                                                \\ \hline
    miniFE.400               & 103.14                                                                                 & 78.18                                                                                \\ \hline
    XSBench.large              & 77.8                                                                                   & 73.56                                                                                \\ \hline    
  \end{tabular}
  \end{center}
  \end{table}

The characterization of our memory system encompasses various metrics, including bandwidth, latency, capacity, power performance, and others. %The interplay between bandwidth and latency is influenced by the specific platform and type of memory employed. For instance, in KNL systems utilizing HBM and DRAM, we might encounter similar latency values but distinct bandwidth levels. %The capacity metric holds significance, particularly due to scenarios where certain applications demand more storage capacity than what smaller memory options can provide. Additionally, differentiation between bandwidth and latency is possible through considerations of read and write operations. Notably, in reference to~\cite{vanrenen}, the performance asymmetry between NVM's write and read functionalities is evident.

%Deriving the precise values for each metric is not always a straightforward task. We are presented with some avenues to acquire these values. For instance, when it comes to metrics like latency and bandwidth, we can query the Heterogeneous Memory Attributes Table (HMAT). While these tables offer vendor-provided information, it is important to acknowledge that older devices might lack such details. Another approach for metric value acquisition involves benchmarking. Nevertheless, it is worth highlighting that the resulting values are contextual to the specific benchmark employed.

%\reviewertwo[]{10. On page 4 you discuss the method for measuring memory power usage by binding the application to the different memory spaces. Please can you comment on how the performance counters treat memory spaces not used but still drawing power?.} DONE

In our study, the power consumption metrics are primarily gathered through hardware counters during application profiling. Given the absence of dedicated memory-kind power consumption counters, we resort to employing the global memory power counter. To differentiate power consumption, we bind the entire process to each memory type, allowing us to work around this limitation. %That is, although the measurement considers the global memory system, i.e., all memories, the binding allows the consumption to correspond to the target memory since the other memories will only have their idle consumption, and this value is removed. %In order to bind an application to a specific memory device, this command was performed, taking into account that \"A\" is the application, \"P\" is the socket, and \"N\" is the NUMA memory:

%\begin{lstlisting}[language=bash,basicstyle=\scriptsize]
%  $ hwloc-bind package:$P.pu:all --membind numa:$N $A
%\end{lstlisting}

In the context of power-limited environments, the power consumption metric assumes vital importance. This is exemplified in~\cite{ivypeng}, where NVM showcases higher power consumption efficiency compared to DRAM. Given our utilization of an HMS featuring three distinct memory types, the power performance ranking's outcome remains less clear-cut.
  
In Table~\ref{tab:powerbeha}, the disparities in power consumption across various applications are apparent (further elaborated in \ref{sec:evaluation}). These variations underline the potential for power savings when opting for NVM over DRAM.

\subsection{Application}
\label{ssec:application}

%\reviewertwo[]{12. In Section 5.2, you use the term "memory-bound". Please be more precise here: are they all bound by memory bandwidth, memory latency, memory capacity, or do they all have a different bound?} WIERD COMMENT

%The interaction dynamics between an application and the memory system are remarkably diverse. This perspective becomes apparent when considering a set of buffers to be allocated, wherein each buffer may manifest unique memory access patterns. %Furthermore, the buffer's size can impose limitations on the choice of memory type, particularly when the memory's capacity proves inadequate. This intricate interplay implies that certain metrics could influence an application's preferences, or those of its primary buffers.

In the scope of our study, we deliberately opted for a variety of memory-bound applications. This selection includes two proxy applications from the Exascale Computing Project (ECP) - namely, miniFE and XSBench. We also incorporated the NASA Advanced Supercomputing (NAS) benchmark suite, encompassing applications such as Integer Sort (IS), Conjugate Gradient (CG), Multi-Grid (MG), Discrete 3D Fourier Transform (FT), and Block Tri-diagonal solver (BT). In addition, we subjected the applications to testing using the High Performance Conjugate Gradient (HPCG).% which bears similarity to NAS CG and showcases irregular memory patterns.

%\begin{table}[]
%  \begin{center}
%  \caption{\label{tab:mlc} Bandwidth \& Latency output from MLC.}
%  \fontsize{6pt}{6pt}\selectfont
%  \begin{tabular}{ccccl}
%  \cline{1-4}
%  \multicolumn{1}{|c|}{\textbf{CPU0}}          & \multicolumn{1}{c|}{\textbf{Local DRAM}} & \multicolumn{1}{c|}{\textbf{Local PMEM}} & \multicolumn{1}{c|}{\textbf{Remote DRAM}} &  \\ \cline{1-4}
%  \multicolumn{1}{|c|}{\textbf{BW{[}GiB/s{]}}} & \multicolumn{1}{c|}{105.91}              & \multicolumn{1}{c|}{36.82}               & \multicolumn{1}{c|}{47.0}                 &  \\ \cline{1-4}
%  \multicolumn{1}{|c|}{\textbf{Lat{[}ns{]}}}   & \multicolumn{1}{c|}{89.4}                & \multicolumn{1}{c|}{308.5}               & \multicolumn{1}{c|}{141.1}                &  \\ \cline{1-4}
%  \multicolumn{1}{l}{}                         & \multicolumn{1}{l}{}                     & \multicolumn{1}{l}{}                     & \multicolumn{1}{l}{}                      & 
%  \end{tabular}
%  \vspace{-3em}
%  \end{center}
%\end{table}

\subsection{Profiling}
\label{ssec:profiling}

In conducting our analysis, it is essential to take into account that the evaluations provide insights into power consumption across the complete memory system. This is due to the inherent limitation of tools in distinguishing between power consumption stemming from DRAM and that originating from NVM. However, by leveraging the capabilities of $\tt{hwloc}$, we are able to allocate the entire process to various memory types, thereby facilitating a differentiation of power consumption among them.

To comprehensively understand the behavior of applications, particularly in terms of bandwidth, total power consumption, and memory system power consumption, we have employed performance counters managed by profilers. Our approach involves using two specific tools:
The Intel Performance Counter Monitor (PCM) and Linux perf. In our study, we have utilized several PCM command-line utilities, including pcm-numa, pcm-mem, and pcm-power which retrieves information related to memory such as accessesm throughput and power respectively~\cite{peng2019}. %  serves as a monitoring utility, providing us with valuable insights into various metrics% such as memory bandwidth and memory power consumption~\cite{peng2019}. 

%\begin{itemize}
%  \item pcm-numa, which retrieves information related to memory accesses.
%  \item pcm-mem, which monitors system read and write memory throughput. It can provide similar information for NVM, depending on the HMS.
%  \item pcm-power, which provides information on the states of the processor and, in our case, can give us power consumption information for the HMS.
%\end{itemize}

The Linux perf tool enables the retrieval of main memory power consumption through the MSR\_DRAM\_ENERGY\_STATUS register. It should be used with $\tt{hwloc-bind}$ to be capable to differentiate the power of each kind of memory. %, an integral component of the Intel Running Average Power Limit (RAPL) interface. %It is worth noting that this register provides valuable insights into memory power consumption. 
%However, it is important to highlight that this particular counter lacks the capability to distinguish between various types of memory connected to the system. For this reason 

This methodology empowers users to not only discern power consumption within memory but also to delineate this consumption based on memory types. The ability to differentiate between them becomes paramount a task achieved in our scenario through power consumption analysis.%, even in the face of limitations imposed by a general memory consumption counter. 
%This distinction is pivotal, as when prioritizing memory hierarchies,
\section{Evaluation}
\label{sec:evaluation}

This section provides a comprehensive overview of the evaluation of our strategy. 
Firstly, we will describe in detail how we exposed the memory system and the emulated kind of memory to increase heterogeneity in our evaluation. Then, we present an analysis of power consumption given an HMS. %, covering various aspects that were considered. 

\begin{figure}[!htb]
    \center{\includegraphics[width=0.6\linewidth]
    {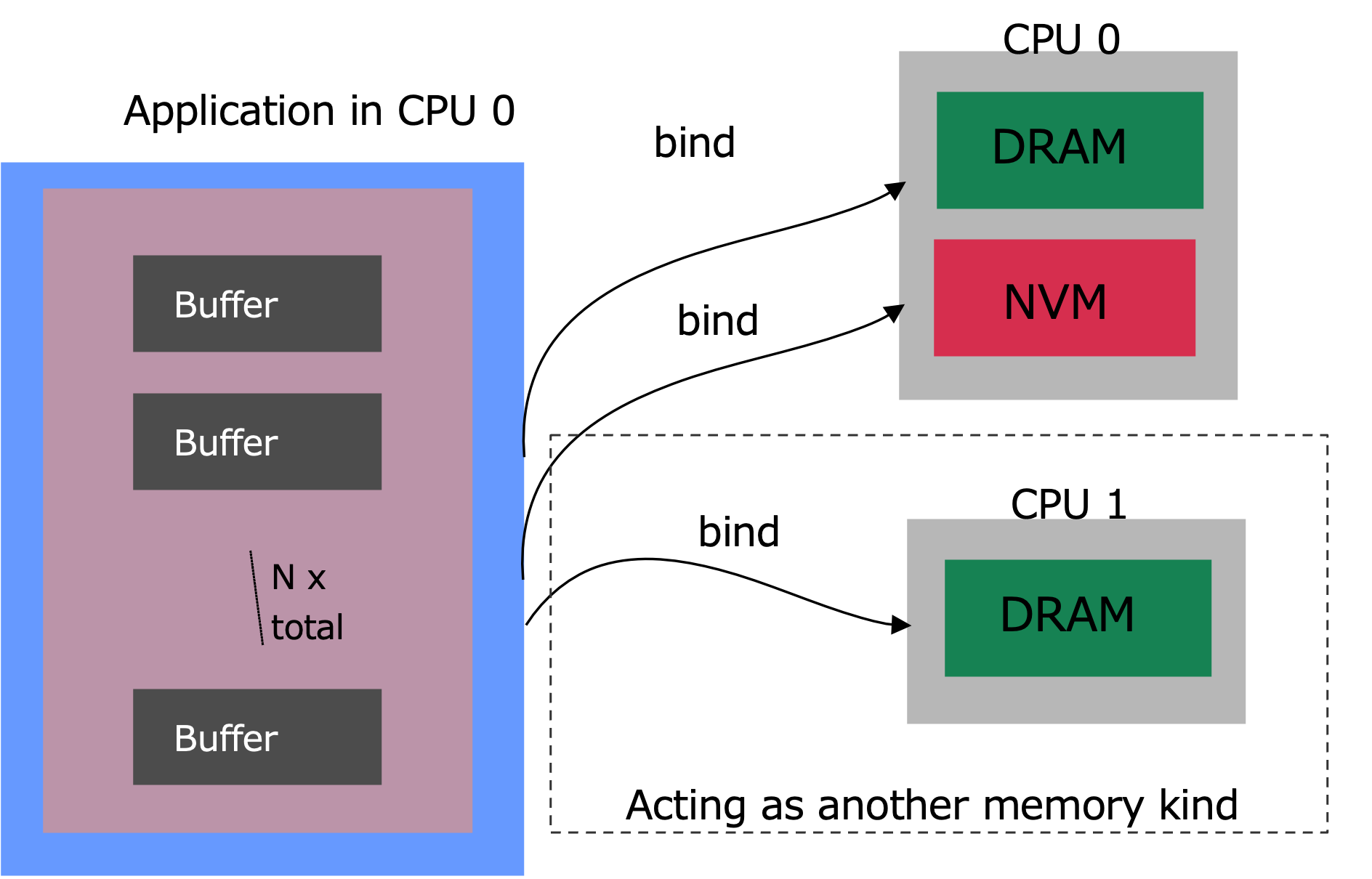}}
    \caption{\label{fig:emulation} Emulation by binding application process to a remote memory target.}
  \end{figure}

\subsection{Emulating Heterogeneous Memory Systems}

Our evaluation encompasses two memory types: DRAM and NVM. To enhance the diversity of our assessment, we have introduced a third memory type through emulation. Through a clever mechanism, we have successfully replicated this third type by leveraging one of CPU0's remote access memories, such as CPU1's local DRAM as indicated in Figure~\ref{fig:emulation}.

%As mentioned previously, the measurement of DRAM and NVM was carried out by means of binding the processes to different numa nodes to make the differentiation through the MSR\_DRAM\_ENERGY\_STATUS counter which presents information per socket. 

%In the case of emulation when using a remote numa node, it should be mentioned that the counter corresponding to the remote socket was used. It is important to clarify that when there are remote accesses the power consumption is normally higher, especially in the counters corresponding to the total consumption in the sockets, but it should be clarified that since this is an emulation that considers a third type of local memory, we have only considered the power counter information of the remote socket.

This strategic approach yields a memory module with distinct attributes compared to CPU0's local memories. %We substantiate this claim through the utilization of the Intel Memory Latency Checker (MLC) – a renowned tool for pinpointing latency and bandwidth within memory systems. As evidenced in Table~\ref{tab:mlc}, notable disparities in bandwidth and latency become apparent. Notably, the versatility of the MLC tool extends beyond NUMA systems, encompassing HMS configurations that expose distinct memory targets as supplementary NUMA nodes.

\subsection{HMS Power consumption analysis}

%During our analysis, it is important to acknowledge that the evaluations present power consumption including the entire memory system. This is due to the limitations of the tools, which do not allow us to differentiate between power consumption specific to DRAM and that specific to NVM. To address this, we utilized the hwloc tool to bind the entire process to various types of memory. This approach enabled us to discern power consumption associated with each type of memory.

%In our study, we introduced multiple variables to our analysis. Firstly, we considered the number of threads as a variable, given that we conducted evaluations under parallel execution scenarios, a common scenario in HPC applications. Additionally, we introduced the problem size as another variable to our analysis. This consideration was motivated by the need to avoid executions that shows excessive utilization of the Last Level Cache (LLC).  

\begin{figure}[!htb]
  \center{\includegraphics[width=0.75\linewidth]
  {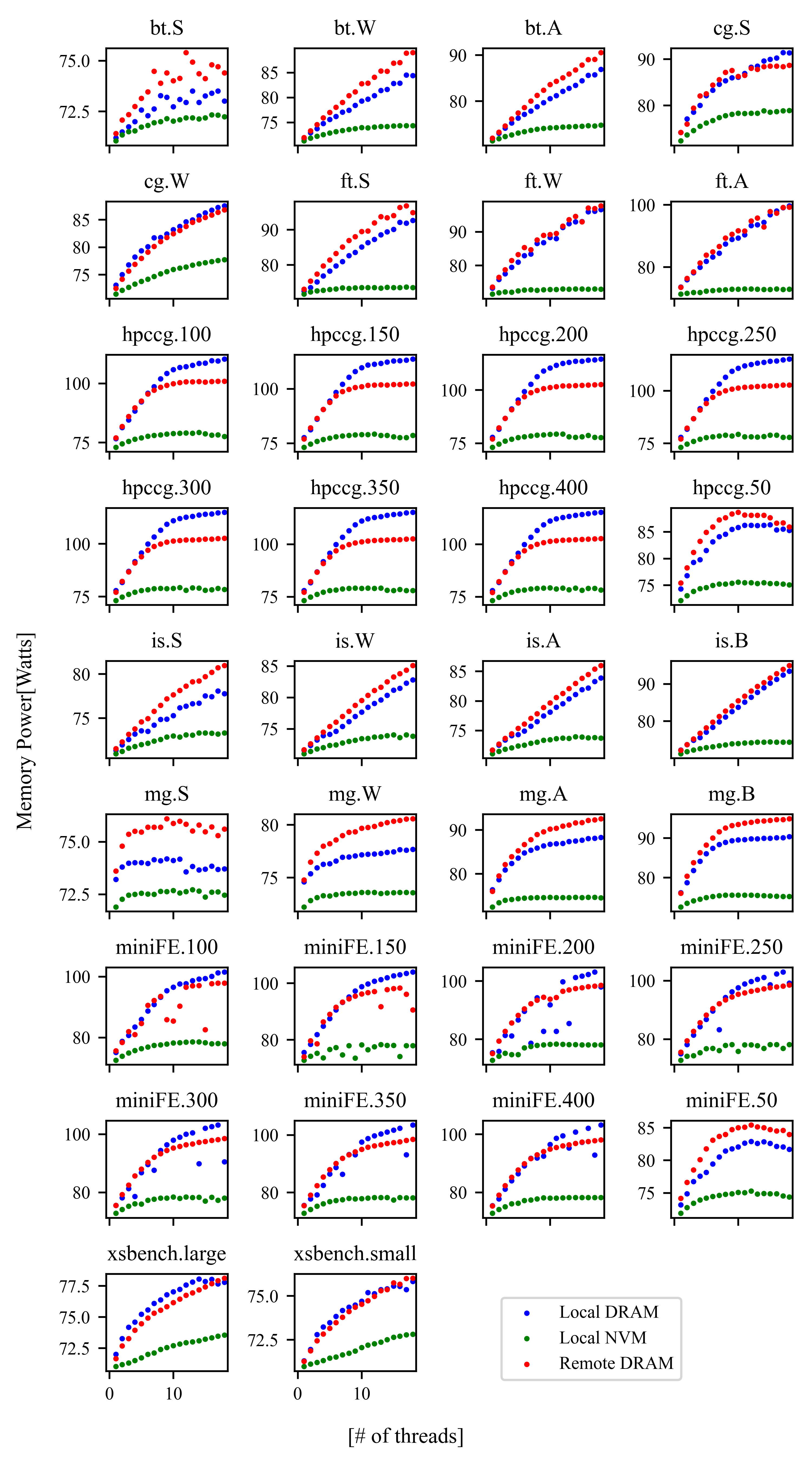}}
  \caption{\label{fig:powerthrds} Memory power consumption behavior taking into account the number of threads on executions bound to Local DRAM, Local NVM, and Remote DRAM.}
  \vspace{-0em}
\end{figure}

In Figure~\ref{fig:powerthrds} we present the power consumption of the Heterogeneous Memory System (HMS) using DRAM and NVM memories and thread counts. Our observations based on this figure include:

\begin{itemize}
  \item In many cases, the power consumption of Remote DRAM is actually lower than that of Local DRAM. To reinforce this notion, it is crucial to reemphasize the concept of simulating memory through the utilization of the NUMA system. %Ordinarily, when remote memory is employed, it involves additional components that contribute to higher power consumption. However, by treating Remote DRAM as another memory kind, the measurement should not encompass the overhead from other components. This perspective shift allows us to more accurately gauge the power consumption of Remote DRAM in isolation. 
  \item  A key characteristic of an HPC application is its utilization of multithreaded executions. While this might result in heightened consumption within components like CPU cores, it is also evident that this phenomenon contributes to an escalation in memory power consumption. This effect arises from the concurrent access to memory facilitated by the multithreading nature of the application.
  \item There is a correlation of memory power consumption when increasing the parallelization level in all kinds of memory, i.e., there is a tendency for having more power consumption when increasing the number of threads. Which memory should be chosen will depend on the application used and also on the memory system. We could say that the generated ordering depends on the memory system and the application. 
  %\item Having three different types of memory within our system, each with its own distinct properties (including simulated memory), it becomes evident that NVM holds a significant power consumption advantage. This positions it at the forefront of the hierarchy and makes it a prime candidate for optimization, particularly in situations where power constraints come into play.
  \item To achieve a equilibrium between power efficiency and performance, the clarity of the situation might not be immediate. It's important to recognize that the second rank in the hierarchy may not consistently correspond to a single memory type. %This intricate interplay emphasizes the crucial role of having a structured methodology, like the one presented, that can assign a ranking to the memory system. %Such an approach empowers both developers and applications to make well-informed decisions, enabling them to select the most suitable memory type capable of striking the desired balance between power efficiency and performance. 

\end{itemize}

%Also in the Figure~\ref{fig:powervsthroughput}, it is noticeable that memory throughput is linked to power consumption, and this connection becomes more pronounced with the use of additional threads. This observation holds significance because metrics like memory throughput could potentially provide users with a relative and indirect insight into power consumption. This could offer valuable insights to users seeking to understand power consumption patterns.

%\begin{figure}[!htb]
%  \center{\includegraphics[width=0.7\linewidth]
%  {img/powervsbwthroughput}}
%  \caption{\label{fig:powervsthroughput} Memory power consumption behavior taking using the biggest problem size vs the memory system throughput generated and by binding to Local DRAM, Local NVM, and Remote DRAM over 1 to 18 threads.}
%  \vspace{-0em}
%\end{figure}

%It is crucial to highlight that this information focuses solely on power consumption performance, disregarding the impact on the overall application performance. Lower power consumption might lead to significant performance degradation. Consequently, this insight holds relevance for developers aiming to strike a balance between speed and power consumption or when the system requires power limited executions.

%\reviewerone[]{6.The paper should discuss some of the fundamental differences in the memory technologies themselves that may be driving the results. It is not clear that the results are so much a product of the heterogeneity of the system or if they are mostly due to these fundamental characteristics, e.g., the need to constantly refresh DRAM.} TODO

\section{Future Work}
\label{sec:futurework}
The exploration of memory system behavior has revolved around determining a ranking that minimizes power consumption among memory types. We acknowledge that ultra-low power consumption can negatively impact application performance. Thus, a more favorable scenario could involve targeting a memory range with lower power consumption, potentially leading to enhanced application performance. This prompts us to delve into comprehending the intricate interplays among various memory metrics. The goal would be to identify optimal trade-offs that developers can leverage when seeking specific balances between these metrics. %For instance, a developer might aim to pinpoint the ideal memory target that strikes a harmonious balance between latency, bandwidth and power consumption.

%As observed, developers lacking access to various HMS face the formidable task of comprehending potential application behavior within such environments. To address this challenge, we are considering the development of predictive models. These models would enable us to anticipate power consumption within our memory system. Additionally, we recognize that developers might encounter situations where predictions for metrics like latency or bandwidth are essential. In terms of future endeavors, we envision extending this approach beyond diverse metrics to encompass a broader array of HMS configurations.

%An additional avenue for extending this research involves not just predicting power consumption within a memory system, but also striving to estimate the overall power consumption. This endeavor would necessitate considering factors like the nature of processing, machine topology, and other relevant variables. Exploring these aspects could potentially offer insights into how various components collectively contribute to power consumption. This broader perspective could yield a more comprehensive understanding of power dynamics within complex computing systems.
\section{Conclusion}
\label{sec:conclusion}

%In this study, our primary focus has been on presenting a strategy in which the different memory systems such as the one we have used can be evaluated in a simple but important manner. 
This approach holds relevance not only for the specific HMS we have utilized, but also for other potential HMS configurations in principle. Of course, the strategy's extension would entail accommodating the intricacies that arise from accessing distinct memory types. %The inherent flexibility of this strategy lends itself well to broader applications and variations in memory architectures.

We have successfully derived an ordering among different memory types for a range of applications. Frequently, this ranking was not readily apparent, especially in cases where the memory type was unfamiliar. 
%This approach empowers applications and users by providing a nuanced understanding of power consumption behavior. Consequently, 
This methodology facilitates the creation of profiles that offer significant power savings, while also enabling the development of profiles that aim to strike a harmonious balance between power consumption and application performance. %It is important to mention that this work has used two types of DRAM and NVM memory and an extra one that has been simulated through remote access. As there are other types of heterogeneous memory systems, e.g. DRAM and MCDRAM or DRAM and CXL.mem the strategy would in principle work as long as each type of memory can be expressed as an extra numa node. In case this cannot be done, the binding of processes to the target memory must be ensured in one way or another and in this way the strategy can be used.

Enhancing the user's comprehension of the memory system significantly promotes program portability across different memory architectures. Moreover, this serves to prevent the memory system from being underutilized or improperly employed.

%\section*{Acknowledgment}

\bibliographystyle{IEEEtran}
%\bibliography{IEEEabrv,bib}
\bibliography{bib}

\end{document}